\documentstyle[12pt]{article}
\begin{document}
\begin{center}
{\large\bf
NAMBU-JONA-LASINIO MODEL IN CURVED SPACETIME WITH MAGNETIC FIELD \\}
\vspace{1cm}
B. Geyer 
\footnote{e-mail:
geyer@ntz.uni-leipzig.d400.de}\\
{\it Institute of Theoretical Physics and Center for Theoretical Sciences}\\
{\it Leipzig University, Augustusplatz 10}\\
{\it D-04109 Leipzig, Germany}\\
L.N. Granda
{\it Departamento de Fisica, Universidad del Valle}\\
{\it A.A. 25360, Cali, Colombia}\\
and\\
S.D. Odintsov
\footnote{e-mail: sergei@ecm.ub.es}\\
{\it Department of Physics and Mathematics,
Tomsk Pedagogical University\\
634041 Tomsk, Russia and}\\
{\it Center for Theoretical Sciences,
 Leipzig University, Augustusplatz 10\\
D-04109 Leipzig, Germany}\\ 

\end{center}
\vspace{1.5cm}

\begin {abstract}
We discuss the phase structure of the NJL model in curved 
spacetime with magnetic
field using $1/N$-expansion and linear curvature approximation. The
effective potential for composite fields $\bar\psi \psi$ is calculated
using the proper-time cut-off in the following cases: a) at non-zero curvature,
 b) at non-zero curvature and non-zero magnetic field, and c) at non-zero
curvature and non-zero covariantly constant gauge field. Chiral symmetry
breaking is studied numerically. We show that the gravitational field
may compensate the effect of the magnetic field what leads to restoration
of chiral symmetry.
\end {abstract}
\vspace{1cm}

\section{Introduction}
The role of the magnetic fields in the models of the inflationary 
Universe has been recently discussed in refs. [1,2]. The realization
 of the fact that strong primordial magnetic fields should be considered
 on equal footing with the strong curvature in the early Universe
 may lead to some new effects in different cosmological applications.
 For example, the combined effect of the gravitational field and 
electromagnetic field can produce a significant increase in the 
number of created particles in the early Universe \cite{ilb}. 
Hence, there appears the motivation to study different field 
theories in the combination of external gravitational and 
electromagnetic fields.
 
In the present letter we investigate the Nambu-Jona-Lasinio (NJL)
 model in curved spacetime with magnetic field. Such a model has 
been often considered in particle physics, as a reliable scenario
 for studying such basic phenomena as chiral symmetry breaking
 and the formation of composite bound states. The phase structure
 of the NJL model in curved spacetime is known to be quite rich 
\cite{ina}. The possibility of curvature induced chiral phase 
transitions may be realized there. From another side, the magnetic
 field usually supports the chiral symmetry breaking. Hence, it 
is interesting to study the behaviour of the chiral symmetry under
 the action of combined magnetic and gravitational field.
Note that such a study may be relevant to particle physics due to the
 fact that the Standard Model may be reformulated as (gauged) NJL model.

We start in Section 2 from the description of the NJL model in curved 
spacetime, and we calculate the effective potential for composite fields in 
1/N expansion and in linear curvature aproximation. Working in proper-time 
cut-off regularization, the possibility of chiral symmetry breaking
 induced by curvature is carefully discussed. Section 3 is devoted to
 the situation when the external magnetic field is also present.
 The effective potential is again evaluated, the corresponding gap
 equation is obtained and
 the chiral symmetry breaking under the action of external gravitational
 and magnetic fields is numerically investigated for some values of the
 coupling constant. In Section 4 we show how to find the effective potential
 for the NJL model in curved spacetime with
 covariantly constant external gauge field.
 
\section{NJL model in curved spacetime: proper-time cut-off}
Let us start from the action for the Nambu-Jona-Lasinio model
 in curved space-time:
\par$$
S= \int d^4 x \sqrt{-g} \left (\bar{\psi}i \gamma^{\mu}(x) \nabla_{\mu} \psi + \frac{\lambda}{2N} \left ((\bar{\psi}\psi)^2 + (\bar{\psi}i\gamma_5 \psi)^2 \right )\right)
\eqno{(1)},$$
where the spinor $\psi$ has N components and $\gamma^{\mu}(x)$ are 
the curved spacetime Dirac matrices.

The standard way to study this model is to work in the scheme of 
the 1/N expansion. We will limit ourselves to the leading order in
 the 1/N expansion. Introducing the auxiliary fields $\sigma(x)$ 
[Band $\pi(x)$ one can rewrite the action (1) in the equivalent form
\par $$
S= \int d^4 x \sqrt{-g} \left (\bar{\psi}i \gamma^{\mu}(x) \nabla_{\mu} \psi 
- \frac{N}{2\lambda}\left(\sigma^2+\pi^2\right)
-\bar{\psi}(\sigma+i\gamma_5\pi)\psi\right).
\eqno{(2)}$$
It is known that even in flat space the global abelian chiral symmetry
 of above model is broken spontaneously when the coupling constant is 
supercritical one. In curved spacetime the situation is getting more 
rich \cite{ina}. There appears the possibility
 of curvature induced phase transitions between symmetric and non 
symmetric phases. In other words, symmetry may be broken even below 
the critical coupling constant (depending on the curvature).

In this section we will discuss the symmetry breaking in the NJL 
model working in the linear curvature aproximation (for a general 
review of the effective action in linear curvature aproximation, 
see \cite{buc}). This model has been already studied in the
 linear curvature approximation (see first paper in \cite{ina}), 
using local momentum representation with ultraviolet cut-off in 
momentum integrals.

We will work in proper-time representation \cite{jul} and will use 
proper-time cut-off. This method will be convenient below for 
generalization to the situation when in addition the external 
magnetic field is present. Integrating over fermionic fields in the
 generating functional for the theory (2) we will get the semiclassical
 effective action
\par $$
S_{eff}=-\int d^4 x \sqrt{-g}\left(\frac{1}{2\lambda}(\sigma^2+\pi^2)
-i\ln\det\left[i\gamma^{\mu}(x)\nabla_{\mu}-(\sigma+i\gamma_5\pi)\right]
\right);
\eqno{(3)} $$
note that N has been factored out.

Let us now discuss the effective potential, where as usually due to
 dependence of V from the invariant $\sigma^2+\pi^2$ it is enough
 to put $\pi=0$. Using the same technique as in the first reference in 
[4] we can show that
\par$$
V(\sigma)=\frac{\sigma^2}{2\lambda}+i\ln\det\lbrack i\gamma^{\mu}(x)
\nabla_{\mu}-\sigma\rbrack.
\eqno {(4)} $$
Then, working in terms of the derivative of V with respect to $\sigma$ 
and using linear curvature expansion for the propagator we get
 (see first paper in \cite{ina})
\par$$
V'(\sigma)=\frac{\sigma}{\lambda}-iTr\int{\frac{d^4k}{(2\pi)^4}
\left[\frac{\hat{k}+\sigma}{k^2-\sigma^2}-\frac{R}{12}\frac{\hat{k}
+\sigma}{(k^2-\sigma^ 2)^2}\quad\right.}
$$
$$
\quad\quad\;\;\left.+\frac{2}{3}R_{\mu\nu}k^{\mu}k^{\nu}\frac{\hat{k}
+\sigma}{(k^2-\sigma^2)^3}
-\frac{1}{2}\gamma^a\sigma^{cd}R_{cda\mu}\frac{k^{\mu}}{(k^2-\sigma^2)^2}\right]
\eqno{(5)}$$
 
\noindent where $\hat{k}=\gamma^{\mu}k_{\mu}$.
After the Wick rotation and calculation of trace we obtain
\par$$
V'(\sigma)=\frac{\sigma}{\lambda}
+\frac{\sigma}{(2\pi)^2}\int_0^{\infty}k^3dk\left[-\frac{1 }{k^2+\sigma^2}
 -\frac{R}{12(k^2+\sigma^2)^2}
+\frac{R}{6}\frac{k^2}{(k^2+\sigma^2)^3}\right],
\eqno{(6)}$$
The expression (6) may be rewritten in the following way
\par $$
V'(\sigma)=\frac{\sigma}{\lambda}+\frac{\sigma}{(2\pi)^2}
\int_{1/\Lambda^2}^{\infty}{ds\left(-\frac{1}{s^2}
+\frac{R}{12s}\right) e^{-s\sigma^2}}.
\eqno{(7)}$$
where the representation
\par$$
\frac{1}{(k^2+\sigma^2)^{\nu}}=\frac{1}{\Gamma(\nu)}
\int_0^{\infty}ds s^{\nu-1}\exp(-s(k^2+\sigma^2))
\eqno{(8)}$$
has been used, integration over $k$ has been done and the
 ultra-violet proper-time cut-off is introduced directly into the integral.

Integrating over $\sigma$ we get
\par $$
V(\sigma)=\frac{\sigma^2}{2\lambda}+\frac{1}{8\pi^2}
\int_{1/\Lambda^2}^{\infty}ds \exp(-s\sigma^2)\left[\frac{1}{s^3}
-\frac{R}{12s^2}\right]
$$
$$
=\frac{\sigma^2}{2\lambda}
+\frac{1}{(4\pi)^2}\left[(\Lambda^4-\Lambda^2\sigma^2)
\exp(-\frac{\sigma^2}{\Lambda^2})\right.
$$
$$
\left.\quad \quad +\sigma^4 Ei(\frac{\sigma^2}{\Lambda^2})
-\frac{R}{6}\left(\Lambda^2 \exp(-\frac{\sigma^2}{\Lambda^2})
-\sigma^2 Ei(\frac{\sigma^2}{\Lambda^2})\right)\right],
\eqno{(9)}
$$ 
where $Ei(x)=\int_x^{\infty}\frac{\exp(-t)}{t}dt=-\gamma-\ln x
-\sum_{n=1}^{\infty}\frac{(-1)^nx^n}{n n!}$ and $\gamma$ is the Euler
 constant.

Expanding Eq. (9) and keeping only terms which are not zero at 
$\Lambda\to\infty$ we get
\par$$
V(\sigma)=\frac{\sigma^2}{\Lambda^2}
-\frac{1}{(4\pi)^2}\left[2\Lambda^2\sigma^2
+\sigma ^4\left(\ln\frac{\sigma^2}{\Lambda^2}+\gamma
-\frac{3}{2}\right)\right.
$$
$$
\left.+\frac{R\sigma^2}{6}\left(\ln\frac{\sigma^2}{\Lambda^2}+\gamma-1\right)
\right]+0(\frac{1}{\Lambda^2})
\eqno{(10)}$$
Thus, we have got the effective potential with proper-time cut-off.

Using Eq.(10) the gap equation is found as follows
\par$$
\frac{4\pi^2}{\lambda\Lambda^2}-1=
\frac{\sigma^2}{\Lambda^2}\left(\ln\frac{\sigma^2}{\Lambda^2}
+\gamma-1\right)+\frac{R}{12\Lambda^2}\left(\ln\frac{\sigma^2}{\Lambda^2}
+\gamma\right).
\eqno{(11)}$$
In flat space-time it is well known that the chiral symmetry is
 broken only for the coupling constant above some $\lambda_0$
\par $$
\lambda\ge\lambda_0=\frac{4\pi^2}{\Lambda^2}.
\eqno{(12)}$$
In curved spacetime, depending on the coupling constant and on the 
curvature, the situation is more interesting \cite{ina}. Particulary, 
for some fixed $\lambda$ one can find the possibility of curvature 
induced first order phase transition (see \cite{ina}). In Fig. 1 we
 have plotted the effective potential $V(\sigma,0)$ for different 
values of R (given in units of $\Lambda^2$ and for the coupling 
$\lambda=1.25\lambda_0)$.

 
The solution of the gap equation (11) gives us the vacuum expectation
 value of the composite field $\bar{\psi}\psi$ and is equal to the 
dynamical mass
of the fermion. In Fig. 2 we have plotted the dynamical mass of the
 fermion field as a function of the curvature $R$ for fixed 
$\lambda=1.25\lambda_0$.

 
\section{NJL model in curved spacetime with magnetic field}
Let us discuss now the situation when the NJL model is considered under the 
action of external gravitational and magnetic fields. The magnetic
 field may be treated exactly in the proper-time method [6] 
(for a review of the proper-time method in external electromagnetic
 fields see [7]). It is quite known that the magnetic field 
increases strongly the possibility of the dynamical symmetry 
breaking (see, for example [8]), if compared with the situation
 without it. The external gravitational field will be taken 
into account in linear curvature aproximation discussed in the
 previous section (this aproximation is perfectly enough to take
 into account the gravitational effects even in GUT epoch). Moreover,
 in linear curvature aproximation we discuss only curvature terms 
which explicitly do not depend on the less important linear 
curvature-magnetic contributions.

With all these remarks, repeating basically the steps of section 2
 (the covariant derivative is now
 $\tilde{\nabla_{\mu}}=\nabla_{\mu}-ieA_{\mu},\; A_{\mu}
=-B x_2 \delta_{\mu 1}$) one can get the following 
effective potential
\par$$
V(\sigma)=\frac{\sigma^2}{2\lambda}
+iTr\ln\left[i\gamma^{\mu}(x)\tilde{\nabla_{\mu}}-\sigma\right]\quad\quad\quad
$$
$$
=\frac{\sigma^2}{2\lambda}
+\frac{1}{8\pi^2}\int_{1/\Lambda^2}^{\infty}
{\frac{ds}{s^2}}\exp(-s\sigma^2)\left[|eB|\coth(s|eB|)
+\left(-{\frac{R}{12}}+0(R^2)\right)\right].
\eqno{(13)}$$
In the absence of the gravitational field ($R=0$), the effective
 potential corresponds to flat space situation where the magnetic 
field is treated exactly [6]. In the absence of the magnatic field
 ($B=0$) we are back to the potential (9). In addition, in 
the linear curvature terms the effect of the magnetic field is not
 taken into account (it will be shown in another place that such terms
 are not relevant compared with the terms written in (13)).

Making the calculation of the integrals in Eq. (13) up to
 $0(1/\Lambda^2)$, and taking the derivative with respect to $\sigma$
 one gets the gap equation
\par$$
\frac{\partial V}{\partial \sigma}=0
\eqno{(14)}$$
as follows
\par$$
\frac{4\pi^2}{\lambda\Lambda^2}-1
=-\frac{\sigma^2}{\Lambda^2}\ln\frac{(\Lambda|eB|^{-1 /2})^2}{2}
+\frac{|eB|}{\Lambda^2}\ln\frac{(\sigma|eB|^{-1/2})^2}{4\pi}+
\gamma\frac{\sigma^2}{\Lambda^2}
$$
$$
+\quad\quad 2\frac{|eB|}{\Lambda^2}\ln\Gamma\left(\frac{\sigma^2|eB|^{-1}}{2}\right)
-\frac{R}{12\Lambda^2}\left(\ln\frac{\Lambda^2}{\sigma^2}-\gamma\right)
+0(\frac{1}{\Lambda}).
\eqno{(15)}
$$
Using this gap equation one can study the dynamical symmetry breaking 
in different cases. In some cases it can be given analytically, for 
example, for values of the coupling constant $\lambda$ much below the 
critical value, i.e.,
\par$$
\lambda\ll\frac{4\pi^2}{\Lambda^2} .
\eqno{(16)}$$
One can find (supposing that the second term in the r.h.s. of (15)
 is the leading one)
\par$$
\frac{4\pi^2}{\lambda\Lambda^2}
-1\approx\frac{|eB|}{\Lambda^2}\ln\frac{(\sigma|eB|^{-1/2 })^2}{4\pi}
-\frac{R}{12\Lambda^2}\ln\frac{\Lambda^2}{\sigma^2}
\eqno{(17)}
$$
and finally, the dynamically generated fermionic mass is given by
\par$$
\sigma^2\approx\left[\left[\frac{|eB|^{-1}}{4\pi}\right]^{-|eB|/\Lambda^2}
(\frac{1}{\Lambda ^2})^{-R/(12\Lambda^2)}
\exp\left(\frac{4\pi^2}{\lambda\Lambda^2}-1\right)
\right]^{1/(|eB|/\Lambda^2+R/( 12\Lambda^2))}
\eqno{(18)}
$$
In the absence of the magnetic field it gives the analytic expression for
the dynamically generated fermionic mass due to the curvature
\par$$
\sigma^2\approx\Lambda^2
\exp\left(\frac{12\Lambda^2}{R}\left[\frac{4\pi^2}{\lambda\Lambda^2}
-1\right]\right) .
\eqno{(19)}$$
One can see that positive curvature tends to make the first term 
in (17) less, i.e. it acts against the dynamical symmetry breaking.
 At the same time, the negative curvature always favors the dynamical
 chiral symmetry breaking in accordance with the explicit calculations
 in external gravitational field [4] and with general considerations of 
ref. [9].

One can give the behaviour of the effective potential as a function of
 curvature and magnetic field for fixed four-fermion coupling constant. 
Fig. 3 shows the breakdown of the symmetry by the magnetic field 
($|eB|/\Lambda^2=0.1$) and the restoration of the
 symmetry by the gravitational field at ($R/\Lambda^2=0.9$), where 
the coupling constant $\lambda$ takes the value
$\lambda=1.25\lambda_0$, where $\lambda_0$ is given in Eq. (12). 
In agreement with Eq. (17), negative values of curvature favor 
 symmetry breaking. Fig. 4 shows the same effect for the coupling
 $\lambda=0.5\lambda_0$. It is seen from Figs. 3 and 4 that
 a  second order phase transition takes place as $R$ changes.
In Fig. 5 we have plotted the solution of the gap equation (15) as a 
function of $R$ for two different values of the coupling $\lambda$
with the same magnetic field.



\section{NJL model in curved spacetime with covariantly constant 
gauge field}
In this section we will discuss the NJL model within covariantly constant
 gauge field and in a weekly curved spacetime. We consider massless NJL 
model of ref. [10] with $SU(2)_R\times SU(2)_L$ chiral symmetry.
The model is described by the Lagrangian 
\par$$
L=\bar{q}i\gamma^{\mu}\bar{\nabla}_{\mu} q+g\left[(\bar{q}q)^2
+(\bar{q}i\gamma_5\tau q)^2\right],
\eqno{(20)}$$
where the external $SU(3)$ gauge field is considered, $q$ is the 
spinor doublet of $SU(3)$ (with components $q^a, a=1,...8$).
Working with covariantly constant $SU(3)$ gauge field expressed 
in terms of invariants $H$ and $E$,
\par$$
H {\rm  resp.} E=\left\{\left[(\frac{1}{4}G_{\mu\nu}^{a\;2})^2
+(\frac{1}{4}G_{\mu\nu}^a\tilde{G}_{\ mu\nu}^a)^2\right]^{1/2}
\pm\frac{1}{4}G_{\mu\nu}^{a\;2}\right\}^{1/2},
\eqno{(21)}$$
and again expressing the effective potential in terms of the auxiliary 
fields (for details in flat space, see [10]) one can get
\par$$
V=\frac{\sigma^2+\pi^2}{4g}
+\frac{N_f}{8\pi^2}tr_c PV\int_{1/\Lambda^2}^{\infty}\frac{ds}{s^3}
e^{-s(\sigma^2+\pi^2)}(g^2EHs^2)\cot(gEs)\coth(gHs)
$$
$$
-i\frac{N_f}{8\pi^2}tr_c\sum_{n=1}^{\infty}
\frac{g^2 HE}{n}\coth(HE^{-1}n\pi)
\exp\left[-\frac{n\pi(\sigma^2+\pi^2)}{gE}\right]
$$
$$
-\frac{N_f}{8\pi^2}\frac{R}{12}tr_c \int_{1/\Lambda^2}^{\infty}
\frac{ds}{s^2}\exp(-s(\sigma^2+\pi^2)),
\eqno{(22)}$$
where $N_f$ corresponds to the flavor symmetry, $tr_c$ means trace 
over color space and $PV$ means the principal value. The same 
cut-off regularization is applied. The imaginary part of the 
effective potential describes the particle creation (for a review, 
see [7]).

Using the real part of the above effective potential one again can
study  numerically the dynamical symmetry breaking under the action 
of covariantly constant gauge field and weak gravitational 
field. The qualitative behaviour of the effective potential in 
chromomagnetic field taking into account the curvature effects is 
similar to the behaviour in the previous section.
We do not discuss this model in detail because the vacuum is not 
stable due to the presence of the imaginary part in an effective potential.

In summary, in this paper we studied the NJL model in curved
 spacetime with magnetic field. Working in linear curvature 
expansion and in the proper-time cut-off, the effective potential 
for the composite field $\bar{\psi}\psi$ has been calculated. The 
phase structure of the theory has been studied numerically first 
for the case without magnetic field. In a situation, when the magnetic 
field is not zero the phase structure has been studied under the 
combined action of magnetic field (which gives stronger effect
 to dynamical symmetry breaking) and of gravitational field. 
Finally, we have shown how one can extend the results of the 
present study to the background of curved spacetime with 
covariantly constant gauge fields.

The approach described in this paper may be extended to the 
situation when both gravitational and magnetic fields are treated 
exactly (with some technical details being more complicated), or 
to other background on gauge fields (say, external electromagnetic
 field).

The results of the above study maybe useful for cosmological 
applications (in particulary, the early Universe) in inflationary 
models based on the composite fields potential of sort (10), (13). 
Note also that it would be of interest to discuss more realistic 
gauged NJL model in a similar way.
\vspace{1cm}\\
\noindent
\bf Acknowledgments\rm\\
SDO would like to thank Yu. I. Shilnov for participation at early 
stage of this work and for independent derivation of results of Section 2 
 and is grateful to D. Amati,
R. Iengo and R. Percacci for kind hospitality at SISSA where this 
work has been completed. This work has been supported by Generalitat 
de Catalonia, Spain and SISSA, Italy.\\
BG would like to thank E. Elizalde for hospitality at Department E.C.M.,
University of Barcelona, where part of the work has been done.

\vspace{2cm}
\begin{figure}[ht]
\caption{The behaviour of the effective potential $v=V/\Lambda^4$ as a function of $x$ ($x=\sigma/\Lambda)$ for different values 
of $r$ ($r=R/\Lambda^2$), for fixing $\lambda=1.25\lambda_0$}
\end{figure}
\begin{figure}[ht]
\caption{The solution of the gap equation (11) ($x=\sigma/\Lambda$ is shown as a function
of the curvature $r$ ($r=R/\Lambda^2$) for $\lambda=1.25\lambda_0$}
\end{figure}
\begin{figure}[ht]
\caption{The behaviour of the effective potential $v=V/\Lambda^4$  for different values of $r$ ($r=R/\Lambda^2$), for fixing $\lambda=1.25\lambda_0$ and magnetic field $|eB|/\Lambda^2=0.1$}
\end{figure}
\begin{figure}[ht]
\caption{The behaviour of the effective potential $v=V/\Lambda^4$  for different values of $r$ ($r=R/\Lambda^2$), for fixing $\lambda=0.5\lambda_0$ and magnetic field $|eB|/\Lambda^2=0.2$}
\end{figure}
\begin{figure}[ht]
\caption{The solution $x=\sigma/\Lambda$ of the gap equation (15) as a function of the curvature $r$ ($r=R/\Lambda^2$) for $\lambda=1.25\lambda_0$ and $\lambda=0.5\lambda_0$}. 
\end{figure}
 
\end{document}